\documentclass{elsart}
\usepackage{graphics}
\begin{document}
\begin{frontmatter}
\hyphenation{Coul-omb ei-gen-val-ue ei-gen-func-tion Ha-mil-to-ni-an
  trans-ver-sal mo-men-tum re-nor-ma-li-zed mas-ses sym-me-tri-za-tion
  dis-cre-ti-za-tion dia-go-na-li-za-tion in-ter-val pro-ba-bi-li-ty
  ha-dro-nic he-li-ci-ty Yu-ka-wa con-si-de-ra-tions spec-tra
  spec-trum cor-res-pond-ing-ly}
\title{On rotations in front-form dynamics} 
\author{Uwe Trittmann$ ^a$  \and 
	Hans-Christian Pauli$\,^b$}
\address{$^a$Department of Physics, 
         Ohio State University, 
	 Columbus, OH, USA \newline
	 $^b$Max-Planck-Institut f\"ur Kernphysik, 
         Heidelberg, Germany }
%
%
\begin{abstract}
Quantum field theories in front-form dynamics are not manifestly 
rotationally invariant. We study a model bound-state equation in 
3+1 dimensional front-form dynamics, which was shown earlier to 
reproduce the Bohr and hyperfine structure of positronium.   
We test this model with regard to its rotational symmetry and 
find that rotational invariance is preserved to a high degree. 
Also, we find and quantify the expected dependence on the cut-off. 
\end{abstract}
\maketitle
\end{frontmatter}

The framework of Discretized Light Cone Quantization (DLCQ) has been
applied successfully to many and diverse physical systems, {\it cf.} 
Ref.~\cite{BrodskyPauliPinsky} and references therein. 
The method is especially efficient when applied in lower
dimensions, and seems ideally suited for strictly 
two-dimensional systems \cite{tHooft,Bassetto}. 
The extension of the DLCQ program into the physical four dimensions 
is both necessary and exciting, 
even when some of its striking advantages are lost on the way.
Several attempts to do so were hampered by a prominent draw-back 
of the front form, particularly
its lack of manifest rotational symmetry;
but recently some progress has been made \cite{Karmanov98}. 

The method of DLCQ was applied to 3+1 dimensions first 
by Tang {\it et al.} \cite{Tang91},
to check the method at the example of positronium. 
This approach, like others \cite{Kaluza92}
trying to solve the matrix eigenvalue problem by using a
light-cone adapted Tamm-Dancoff procedure, 
suffered severe convergence problems.
Wilson and collaborators \cite{Wilson}
considered the problem in more abstract terms, 
emphasizing the role of 
renormalization problems.
But one may state in all fairness that concrete and practical
prescriptions have not emanated thus far.
Their positronium spectrum \cite{Jones96a} was obtained later.
Within the model of Krautg\"art\-ner, Pauli and W\"olz
\cite{KPW92}, however, the Bohr and the hyperfine structure 
of positronium on the light-front was resolved,
but only for the z-component of total angular momentum 
$J_z=0$.
The derivation of the effective interaction was not 
overwhelmingly convincing, 
and the problems with a cutoff-dependent 
eigenvalue equation have not been settled thus far, 
see however also Ref.~\cite{Pau00}.

But even with the apparent shortcomings of this model,
one can examine its rotational symmetry
by asking whe\-ther the corresponding members of a rotational 
multiplet are degenerate or not, and if the multiplets contain the 
correct number of degenerate states. 
The agenda is then quite obvious: 
\begin{itemize}
\item Generalize the method to arbitrary $J_z$;
\item Increase the numerical accuracy needed for that.
\end{itemize}
If we can calculate the eigenvalue spectrum separately for each $J_z$,
which is a kinematic operator both in the 
front and in the instant form \cite{Leutwyler},
some of the eigenvalues of distinct $J_z$ 
must be degenerate and form a multiplet, also in the front form. 
We can thus investigate {\em quantitatively} to which extent  
rotational symmetry is violated within a specific model.
Any violation of the degeneracy of multiplets is a strong indication
that the original covariance of the Lagrangian was lost 
by nature of the (model dependent) approximations.

\section{Model positronium}
\label{Section:model}

The model of positronium considered here has been
introduced by Krautg\"art\-ner {\it et al.} \cite{KPW92}  
and we refer to it for all unquoted details.
In light-cone quantization, the contraction of the momentum operators 
$P^{\mu}$, is called the light-cone
Hamiltonian, $H_{\rm LC}=P^{\mu}P_{\mu}$. Solving the eigenvalue problem 
\begin{equation}
    H_{\rm LC}|\Psi\rangle=M^2|\Psi\rangle
,\label{EVP}\end{equation}
yields the mass (squared) eigenvalue spectrum 
of a physical system.
This full problem is very difficult to solve,
particularly in gauge theory, see \cite{Hil00}.
It is easier to solve the problem in a reduced space
with an effective Hamiltonian \cite{MorseFeshbach} 
\begin{equation}
    H^{\rm eff}_{\rm LC}|\psi\rangle =
    (T + U^{\rm eff}) |\psi\rangle = 
    M^2|\psi\rangle
,\label{effIntEqn}\end{equation}
with some free kinetic part $T$
and an effective interaction $U^{\rm eff}$.
The reduced space is here the Fock space of the
single electron and a single positron, {\it i.e.}
\( |\psi\rangle \equiv |\Psi_{e\bar e}\rangle \).
In what follows, we denote  
the mass, the longitudinal momentum fraction, the 
transverse momentum, and the helicity of the electron 
with $m,x,\vec k _{\!\perp}$, and $\lambda_{e}$, respectively,
and those of the positron
with $m,1-x,-\vec k _{\!\perp}$, and $\lambda_{\bar e}$.
Physically, the effective interaction scatters 
an electron-positron pair
from a state with four-momenta $(k_e,k_{\bar{e}})$ 
into a state with $(k'_e,k'_{\bar{e}})$, 
which in general has a different free invariant mass
than the entrance channel. 
This requires a certain amount 
of symmetrization which was made in all previous work 
\cite{TamDancoff,BLepage80}, and thus also in \cite{KPW92}. 
A more thorough discussion can be found Ref. \cite{TrittmannPauli97}.
This in mind, Eq.~(\ref{effIntEqn}) becomes an integral equation 
\begin{eqnarray} 
    &&M^2_n\langle x,\vec k_{\!\perp}; \lambda_{e},
    \lambda_{\bar e}  \vert \psi_n\rangle =
    {m^{\,2} + \vec k_{\!\perp}^{\,2}\over x(1-x)}\, 
    \langle x,\vec k_{\!\perp}; \lambda_{e},
    \lambda_{\bar e}  \vert \psi_n\rangle
\label{IntEqn}\\ &&\quad\quad
    -{\alpha \over 2\pi^2}
    \sum _{\lambda_e^\prime,\lambda_{\bar e}^\prime}
    \!\int_D {dx^\prime d^2 \vec k_{\!\perp}^\prime
    \over Q^2}\,
    \langle 
       x,\vec k_{\!\perp};\lambda_e,\lambda_{\bar e}
       \vert S\vert 
       x^\prime,\vec k_{\!\perp}^\prime;
       \lambda_e^\prime,\lambda_{\bar e}^\prime
    \rangle
	\langle x^\prime,\vec k_{\!\perp}^\prime; 
    \lambda_e^\prime,\lambda_{\bar e}^\prime  
    \vert \psi_n\rangle
.\nonumber\end {eqnarray}
Symmetrization is reflected in the denominator 
\begin{equation}
   Q ^2 (x,\vec k_{\!\perp};x',\vec k_{\!\perp}') 
   = -\frac{1}{2}\left[(k_{e}-k_{e}')^2 + (k_{\bar e}-k_{\bar e}')^2\right]
,\end{equation}
which is the mean Feynman four-momentum transfer. 
The Lorentz-contracted Dirac spinors are collected in 
\begin{equation}
   \langle\lambda_{e},\lambda_{\bar e}\vert S\vert
   \lambda_{e}^\prime,\lambda_{\bar e}^\prime\rangle =
   \left[ \overline u (k_{e},\lambda_{e})\gamma^\mu
   u(k_{e}^\prime,\lambda_{e}^\prime)\right] \, 
   \left[ \overline v(k_{\bar e}^\prime,\lambda_{\bar e}^\prime) 
   \gamma_\mu 
    v(k_{\bar e},\lambda_{\bar e})\right] 
.\label{teq:4}\end{equation}
In helicity space it is a $4\times 4$ matrix which is 
tabulated explicitly in the Compendium \cite{Com00}
or in Ref.~\cite[App.~B]{TrittmannPauli97b}. 
The domain of integration $D$ is specified by the 
`sharp' cut-off $\Lambda$ 
\begin{equation}
   m^2 +\vec k_{\!\perp}^{'\,2}\leq x'(1-x')(\Lambda^2 +4m^2)
,\end{equation}
{\it i.e.} by Lepage-Brodsky regularization \cite{BLepage80}.
The experimental literature differentiates between 
proper positronium ($\mu\bar e$) and 
true positronium ($e\bar e$) (Telegdi).
The model in Eq.(\ref{IntEqn}) is neither of the two:
It is `model positronium'. 
The annihilation channel is however included in
Refs.~\cite{TrittmannPauli97,TrittmannPauli97b,Trittmann97}.

In the meantime, the inherent difficulties with the
light-cone adapted Tamm-Dancoff approach
have faded away, since better formalisms
yield the very same  integral equation (\ref{IntEqn}), 
see {\it f.e.} \cite{Pau00}.
Note that the above equation is actually a set of four coupled integral
equations in the three variables $x$, 
$k_x=\vert\vec k_{\!\perp}\vert \sin{\varphi}$ and 
$k_y=\vert\vec k_{\!\perp}\vert \cos{\varphi}$.
Its numerical solution is highly non-trivial.

\section{The construction of the multiplets and the numerical results}
\label{Section:Numerics}
 
From usual (instant form) quantization one is used to the fact that
the Hamiltonian commutes with the total angular momentum
operator $\mathbf{J} ^2$ with eigenvalues $J(J+1)$.
As a consequence, the 
Hamiltonian eigenvalues are $(2J+1)$-fold degenerate. 
The individual members of a $\mathbf{J} ^2$-multiplet 
are labeled by $J_z$.
In the front form things are different:
$J _z$ is the only kinematic of the three rotation operators. 
The key observation here is that 
front-form eigenstates can be classified by $J_z$ but not by a 
quantum number with respect to $\mathbf{J}^2$.
We can, however, diagonalize the Hamiltonian subsequently in different 
sectors characterized by $J_z$, {\it i.e.} as a function
of $J_z = 0, \pm 1, \pm 2, \ldots$. 
Some of the eigenvalues will be degenerate in $J_z$,
within a certain numerical accuracy of course, and
form the usual (instant form) multiplets.
The largest value of $J_z$ in a given multiplet determines the multiplicity,
$2 (J_z)_{\rm max} + 1$, and thus indirectly the quantum number $J$. 
If we find these degenerate multiplets, we have reproduced
the multiplet structure of the instant form.
The question is then to which extent multiplets are degenerate, 
{\it i.e.} on which level of accuracy rotational invariance is violated.

\begin{figure}
  \resizebox{0.99\textwidth}{!}{\includegraphics{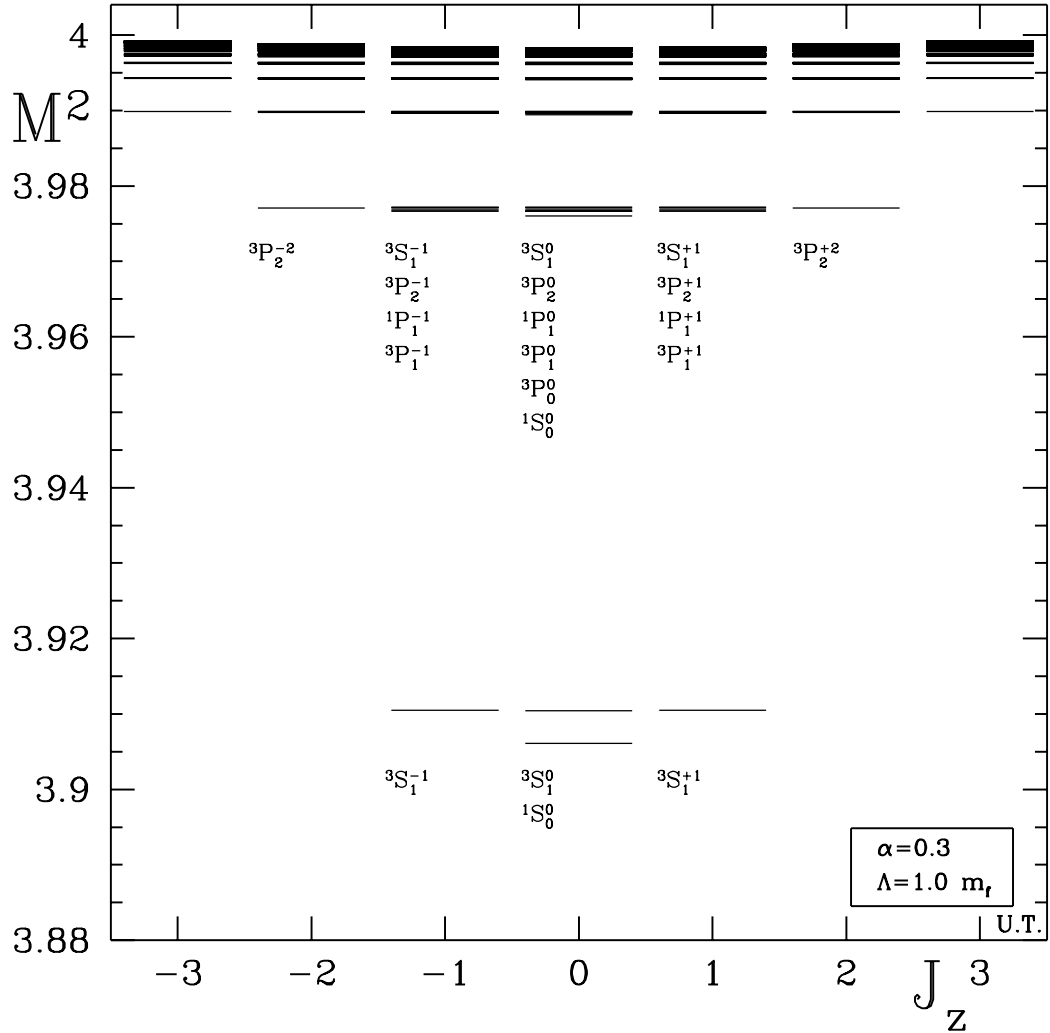}}
\caption{\label{yrast}
   The spectrum of the positronium model for $\vert J_z\vert \leq 3$
   at $\Lambda=m$, $N_1{=}N_2{=}21$.
   The eigenvalues $M$ are given in units of the electron mass $m$.~--
   The multiplet structure of the eigenvalues is emphasized by
   the phenomenological notation $^{2S+1}L^{J_z}_J$.}
\end{figure}

Our prejudices are illustrated in Fig.~\ref{yrast}.
Here is our most important result, 
and we shall explain below how to get it.
The figure has the anticipated properties: 
the eigenvalues are arranged in degenerate multiplets,
and each multiplet has an odd number of members. 
The eigenvalues can be arranged in clusters which are 
characterized by the Bohr quantum number $n$. 
For a pure Coulomb spectrum there would be $(4n-2)$ 
degenerate states with spin projection $J_z=0$, and we 
found that the multiplets in our calculations
pass this check: they have precisely $(4n-2)$ members.
We have observed this property up to $n{=}5$, 
beyond which we have seen no reason to pursue.

Another important result is that only those combinations of the
quantum numbers $\pi_\mathcal{C}$ and $\pi_\mathcal{H}$ appear 
which are expected from a non-relativistic analysis.
Usually one classifies the states with the 
spectroscopic notation $^{2S+1} L_J$.
In front-form dynamics, however, neither the total angular momentum $J$, 
nor the orbital angular momentum $L$, nor the total spin $S$
are good (kinematic) quantum numbers.
Rather, the front-form Hamiltonian is symmetric under charge 
conjugation $\mathcal{C}$ and 
a combination of time reversal $\mathcal{T}$
and parity $\mathcal{P}$, sometimes called {\em handedness} 
$\mathcal{H}\equiv \mathcal{PT}$.
In the non-relativistic case, 
one can relate the (instant form) quantum numbers $(J,L,S)$ 
uniquely to the (front form) quantum numbers 
$(\pi_\mathcal{C}$,$\pi_\mathcal{H})$,
if one chooses a certain convention 
for the time reversal operation \cite{Hornbostel}. 
The spectroscopic notation is used only to
label the states conveniently  
\cite{TrittmannPauli97,TrittmannPauli97b,Trittmann97}.

\begin{table}
\caption{\label{tab:2}
   The binding coefficients $B_n$ 
   for $\alpha=0.3,$ $\Lambda=m$, $N_1=N_2=21$.~---
   Column~1 gives the spectroscopic notation $n\ ^{2S+1} L _J$ and
   column~2 the parities $\mathcal{C}$ and $\mathcal{H}$.
   The numerical errors of $B_n$ (in parenthesis) 
   are estimated, see text. 
   The last column lists the discrepancy to perturbation theory 
   up to order $\mathcal{O}(\alpha^4)$ \protect{\cite{Gupta89}}.} 
\begin{tabular}
{@{}cc@{}c|@{\hskip2em}l@{\hskip2em}l@{\hskip2em}|rr@{}}
\hline\hline
Term & $\pi_\mathcal{C}$ 
     & $\pi_\mathcal{H}$ 
     & $B_n(0)$ & $B_n(1)$ 
     & $\left( B_n(0)-B_n(1)\right) $ 
     & $\Delta B_n(\mathrm {pert})$\rule[-3mm]{0mm}{8mm} \\ 
     & & & & & $\times 10 ^5$ & in \%    \\ \hline
  1~$^1\!S_0$ &$+$&$-$& 1.04955(2)  &     { }     &   { }  & -6.13\\  
  1~$^3S_1$ &$-$&$+$& 1.00101(11) & 1.00038(7)  &  63.5  & -0.29\\
  2~$^1S_0$ &$+$&$-$& 0.26024(17) &     { }     &   { }  &  3.13\\  
  2~$^3S_1$ &$-$&$+$& 0.25380(22) & 0.25372(21) &   8.33 & -0.07\\
  2~$^1P_1$ &$-$&$-$& 0.25797(16) & 0.25798(17) &  -1.30 & -1.71\\ 
  2~$^3P_0$ &$+$&$+$& 0.26707(16) &     { }     &   { }  & -2.27\\ 
  2~$^3P_1$ &$+$&$-$& 0.25967(21) & 0.26008(16) & -40.8  & -1.63\\
  2~$^3P_2$ &$+$&$+$& 0.25526(18) & 0.25525(17) &   0.47 & -1.69\\  
  3~$^1S_0$ &$+$&$-$& 0.11521(31) &     { }     &   { }  &  1.54\\  
  3~$^3S_1$ &$-$&$+$& 0.11344(36) & 0.11341(26) &   2.79 & -0.77\\
  3~$^1P_1$ &$-$&$-$& 0.11449(27) & 0.11453(28) &  -3.96 & -1.71\\
  3~$^3P_0$ &$+$&$+$& 0.11713(27) &     { }     &   { }  & -2.04\\
  3~$^3P_1$ &$+$&$-$& 0.11513(33) & 0.11512(27) &   1.13 & -1.77\\
  3~$^3P_2$ &$+$&$+$& 0.11372(28) & 0.11372(28) &  -0.26 & -1.72\\
  3~$^1D_2$ &$+$&$-$& 0.11282(15) & 0.11284(16) &  -2.66 & -1.02\\
  3~$^3D_1$ &$-$&$+$& 0.11343(16) & 0.11350(28) &  -6.90 & -1.56\\
  3~$^3D_2$ &$-$&$-$& 0.11298(16) & 0.11298(16) &  -0.43 & -1.06\\
  3~$^3D_3$ &$-$&$+$& 0.11251(16) & 0.11252(16) &  -0.41 & -1.03\\
\hline\hline
\end{tabular}
\end{table}

The multiplets of mass (squared) eigenvalues are 
seemingly degenerate on the scale of the figure 
(Fig.~\ref{yrast}).
To be more quantitative we collect in Table~~\ref{tab:2}
the first 18 eigenvalues in the form of
binding coefficient defined by
\begin{equation}
   B_n(J_z) = \frac{4}{m\alpha^2}\ \left(2m-M_n\right) 
,\label{eq:5}\end{equation}
where the trivial dependence on mass and coupling constant 
is removed.
The numerical errors (in parenthesis) are estimated from the 
difference between the values for maximal and next to maximal number 
of integration points. 
Except for a few states to be discussed below,
the relative discrepancy of corresponding eigenvalues 
is typically a few parts in $10^5$, 
{\it i.e.} the degeneracy is on the level of
numerical errors of the diagonalization routine.

The positronium spectrum
has been calculated perturbatively, 
long ago by \cite{Fermi30} and \cite{Gupta89}, 
and we have to keep in mind that these calculations were done 
under the {\it proviso} of the small $\alpha\sim 1/137$.
Despite that, our eigenvalues agree quite well, 
{\it i.e.} on the percent level, 
with those results, even for the highly excited states,
see Table~\ref{tab:2}.

The singlets $^1\!S_0$ tend to be 
too weakly bound, especially for the low Bohr quantum 
numbers $n=1$ and $n=2$. 
This effect is reversed for the triplets $^3\!S_1$.
Also, the ordering of the multiplets seems to have minor errors. 
For instance, the $2\,^1\!S_0$ state and the $2\,^1\!P_0$ state 
are permuted: 
the $S$-state should be the lowest according to perturbation theory.
We convinced ourselves that this is a finite cut-off effect. 

\begin{table}
\caption{\label{tab:1}
  The binding coefficients of the singlet $(B_s)$ and 
  the triplet states 
  $(B_t)$ for $\alpha=0.3,$ $N_1=25, N_2=21$ 
  are given as function of the cut-off $\Lambda$. 
  They are compared with the results of perturbation theory 
  \protect{\cite{Gupta89}} up to order $\mathcal{O}(\alpha^4)$ 
  and $\mathcal{O}(\alpha^6\ln\alpha)$. }
\centerline{
\begin{tabular}
{@{}r@{\hskip3em}|@{\hskip5em}c@{\hskip4em}c@{\hskip5em}c@{}}
\hline\hline
   cut-off $\displaystyle \frac{\Lambda}{m}$ 
             &     $B_s$   &   $B_t$    & $C_{hf}$  \rule[-3mm]{0mm}{8mm} \\ 
   \hline   
   1.0       &  1.04903964 & 1.00046227 &  0.13493713 \\ 
   1.8       &  1.16373904 & 1.06860934 &  0.26424917 \\  
   3.6       &  1.25570148 & 1.10111328 &  0.42941166 \\  
   5.4       &  1.29978050 & 1.11163578 &  0.52262422 \\  
   7.2       &  1.32941912 & 1.11782782 &  0.58775360 \\  
   9.0       &  1.35223982 & 1.12233652 &  0.63862028 \\  
  10.8       &  1.37112216 & 1.12596311 &  0.68099735 \\  
  12.6       &  1.38744792 & 1.12904455 &  0.71778713 \\  
  14.4       &  1.40198469 & 1.13175363 &  0.75064183 \\  
  16.2       &  1.41520247 & 1.13419048 &  0.78058886 \\  
  18.0       &  1.42740143 & 1.13641774 &  0.80828803 \\ \hline 
$\mathcal{O}(\alpha^4)$ & 1.11812500 & 0.99812500 & 0.33333333\\ 
$\mathcal{O}(\alpha^6\ln\alpha)$ & & & 0.23792985\\ 
\hline\hline\end{tabular}}
\end{table}

The numerical approach has two formal parameters:
the number of integration points $N_1=N_2=N$ and $\Lambda$.
The dependence on the number of integration points $N$
was found to decrease exponentially fast, 
even for the splitting between triplet states.
The asymptotic value $\Delta M^2 (N{\rightarrow}\infty)= a$ 
amounts to only 0.5\% of the relevant scale, namely the singlet-triplet 
splitting, of roughly $0.0102m^2$. With other words, the states are 
highly degenerate.

\textbf{The dependence on the cut-off $\mathbf{\Lambda}$}.
The weak point of the present work is the 
expected dependence on $\Lambda$.
Its occurrence is not a particular feature of the
light cone approach, but appears due to the 
Dirac interaction in any (gauge) field theory:
For very large (transversal) momenta, {\it i.e.} for
$\vec k_{\!\perp}^{\prime\,2} \gg \vec k_{\!\perp} ^{\,2}$,
holds  
$\langle\uparrow\downarrow\vert S\vert\uparrow\downarrow\rangle
 /Q^2 \longrightarrow 2$ and 
$\langle\downarrow\uparrow\vert S\vert\downarrow\uparrow\rangle 
 /Q^2 \longrightarrow 2$:
The kernel does not decay sufficiently fast, 
see also Ref.~\cite{Pau00}.

For discussing the dependence on $\Lambda$ to some detail
we present in Table~\ref{tab:1} the binding coefficient
for the singlet and the triplet together with the 
hyperfine coefficient 
\begin{equation}
   C_{hf}=\frac{1}{m\alpha^4}\left(M_{triplet}-M_{singlet}\right)
.\label{eq:6}\end{equation}
Its perturbative value \cite{Fermi30,Gupta89} is known:
\begin{equation}
   C_{hf} = \frac{1}{3}+\left[\frac{1}{4}\right]-
   \frac{\alpha}{2\pi}\left(\ln 2 +\frac{16}{9}\right)
   +\mathcal{O}(\alpha^2\ln\alpha)
.\label{eq:7}\end{equation}
The term in square brackets is the contribution from the 
one-photon annihilation.
When comparing to previous results \cite[Table V]{KPW92},
we find that the singlet falls off faster with $\Lambda$, yielding a
smaller value for the hyperfine coefficient in our calculations.
We note that the predictions of perturbation theory
for different orders differ as much as $40\%$ for the 
value of $\alpha=0.3$ used in the present work.

\begin{figure} [t]  
\begin{minipage}[t]{68mm}
  \resizebox{1.00\textwidth}{!}{\includegraphics{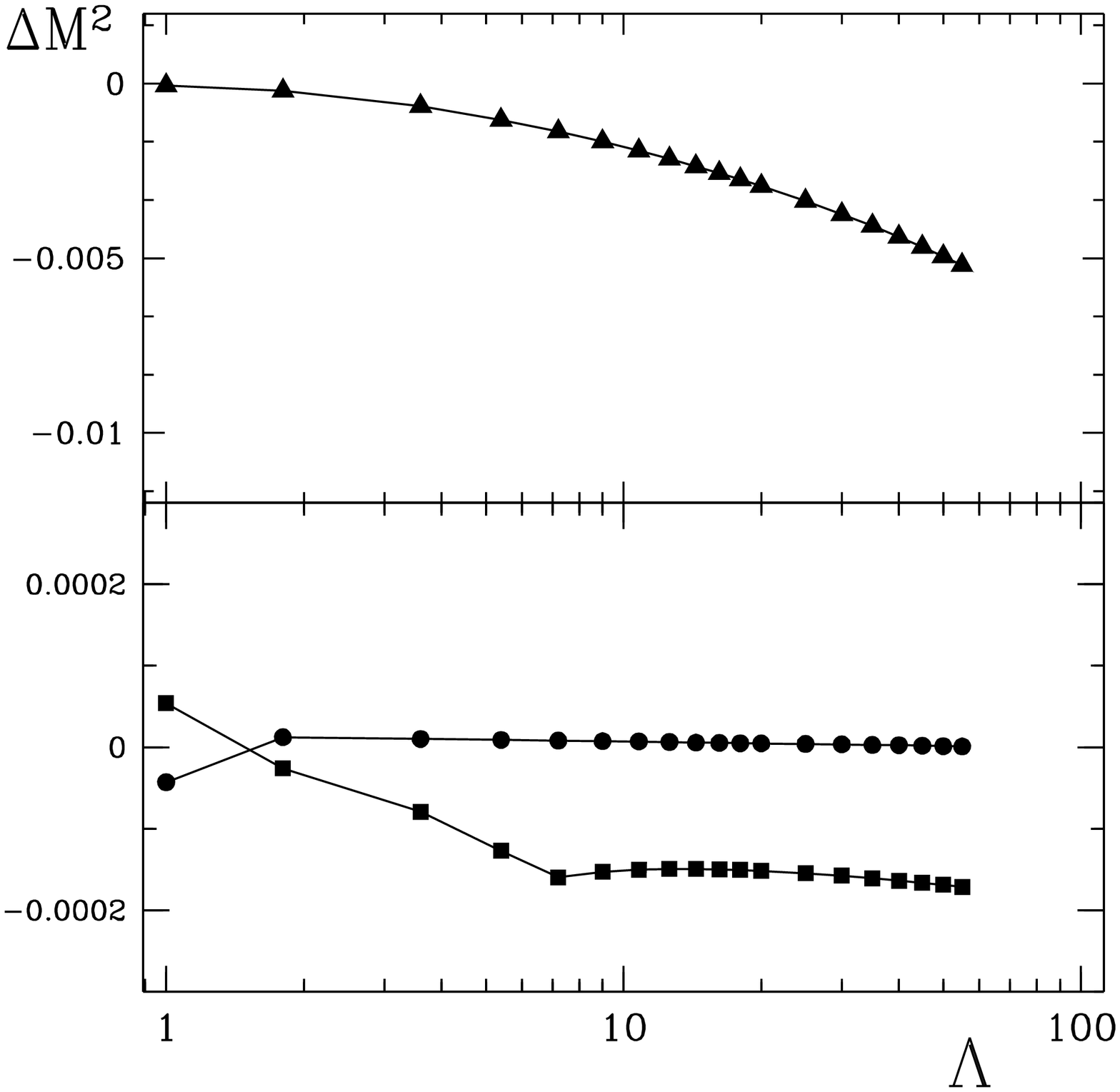}}
\caption{\label{diff} 
   The discrepancy function $\Delta M_i^2$ $(=M_i^2(1)-M_i^2(0))$,
   is plotted versus the cut-off $\Lambda$
   for the $1^3S_1\,(\triangle)$ [upper plot], 
   the $2^3P_1$, and the $2^1P_1\,(\bullet)$ [lower plot].
   Parameters are $\Lambda=m$, $N_1=21$, and $N_2=21$.\hfill}
\end{minipage} \hfill
\begin{minipage}[t]{68mm}
  \resizebox{1.00\textwidth}{!}{\includegraphics{Jz1_fig11.epsi}}
\caption{\label{fig:diff-b} 
   The decrease of the $J_z=1$ triplet ground state wave function 
   with parallel helicities  as a function of the off-shell mass $\mu$. 
   The parameters are $\Lambda=20.0\, m$, $N_1{=}41$, $N_2{=}11$. 
   The six different curves correspond to six values of the 
   angle $\theta$. 
   Rotational symmetry is only for small $\mu$.}
\end{minipage} 
\end{figure}

The degeneracy of eigenvalues as a function of the cut-off $\Lambda$
is displayed in Fig.~\ref{diff}.
One notices that the discrepancy between the triplet ground states 
rises from close to zero at $\Lambda=m$ to roughly 50\% of the 
hyperfine splitting for the large value of $\Lambda=50m$. 
For the excited states the difference stays always below 12\% 
on the relevant scale.  
Since we expect a logarithmic divergence 
we fit the curves $M^2(\Lambda)$ to a polynomial in $\ln \Lambda$.
If one omits the points for $\Lambda>20\,m$ because the 
integrations of the numerical counter term can become problematic 
in this region, the fit yields
\begin{eqnarray}
   M^2_{singlet}(\Lambda)&=& \left(
   3.90545 - 0.03510 \ln{\frac{\Lambda}{m}}  
           + 0.00746 \ln^2{\frac{\Lambda}{m}} \right) m^2
,\nonumber\\
   M^2_{triplet}(\Lambda)&=& \left(
   3.90976 - 0.01858 \ln{\frac{\Lambda}{m}}  
           + 0.00789\ln^2{\frac{\Lambda}{m}}\right) m^2
.\label{LambdaFit} \end{eqnarray} 
The small coefficient of the $\ln^2{\Lambda/m}$-term hints 
thus indeed at a logarithmic cut-off dependence of the eigenvalues.
The dependence weakens if one includes the 
annihilation channel \cite{Trittmann97}.

\section{More details for general $J_z$ and the wavefunction}
\label{sec:HJz}

We want to calculate the Hamiltonian spectrum in all sectors of $J_z$.
It is advantageous to reformulate the eigenvalue problem 
in such a way that allows for a separation of the sectors, {\it i.e.} a 
block diagonalization of the Hamiltonian with respect to the 
conserved quantum number $J_z$.
By exploiting the symmetry of the Lagrangian under rotations in the 
$(x,y)$ plane, we can substitute the angular variable $\varphi$ by the 
quantum number $J_z$ applying a integral transformation on 
Eq.(\ref{IntEqn}), and in particular on the effective matrix elements
\begin{eqnarray}
   &&\langle x,{k}_{\perp}, J_z;\lambda_e,\lambda_{\bar{e}}|
   \widetilde U^\mathrm{eff}|
   x',{k}'_{\perp}, J_z;\lambda'_e,\lambda'_{\bar{e}}\rangle 
\label{PhiInt}\\ &&=
   \frac{1}{2\pi}\int_0^{2\pi} \int_0^{2\pi} d\varphi\; d\varphi'\; 
   e^{-i (L_z \varphi- L_z' \varphi')}
   \langle x,k_{\!\perp},\varphi; 
   \lambda_e,\lambda_{\bar{e}}| U^\mathrm{eff}|
   x',k_{\!\perp}',\varphi'; \lambda'_e,\lambda'_{\bar{e}}\rangle
.\nonumber\end{eqnarray}
It is easy to convince one-self that angular dependence can enter 
only in the form of the difference $\varphi-\varphi'$,
and that the matrix elements 
are well-behaved for an arbitrary $L_z{=}J_z{-}S_z$.
The eigenvalue problem, Eq.~(\ref{IntEqn}), looks now like
\begin{eqnarray}  
    0&=&\left( M^2_n-
    {m^{\,2} + \vec k_{\!\perp}^{\,2}\over x(1-x)}\right) 
    \langle x, k_{\!\perp}; \lambda_{e},
    \lambda_{\bar e}; J_z  \vert \psi_n\rangle 
\label{IntEqnJz}\\ 
	&&+
    \sum _{\lambda_e^\prime,\lambda_{\bar e}^\prime}
    \!\int_D dx^\prime d k_{\!\perp}^\prime
    \langle 
       x, k_{\!\perp};\lambda_e,\lambda_{\bar e}
       \vert \widetilde U^\mathrm{eff}\vert 
       x^\prime, k_{\!\perp}^\prime;
       \lambda_e^\prime,\lambda_{\bar e}^\prime
    \rangle
    \,\langle x^\prime, k_{\!\perp}^\prime; 
    \lambda_e^\prime,\lambda_{\bar e}^\prime; J_z  
    \vert \psi_n\rangle
.\nonumber\end{eqnarray}
The angle averaged `general helicity table' 
$\langle\lambda_e,\lambda_{\bar{e}}\vert\widetilde U^\mathrm{eff}
   \vert\lambda'_e,\lambda'_{\bar{e}}\rangle$
is again a $4\times 4$-matrix in helicity space and tabulated in
Refs.~\cite{TrittmannPauli97,TrittmannPauli97b,Trittmann97}.
 
In the pioneering work of 
Krautg\"artner {\it et al.} \cite{KPW92}
Eq.~(\ref{IntEqnJz}) was solved numerically 
for the special case $J_z=0$. 
By convenience they use momentum coordinates
$(\mu,\theta)$ instead of $(x,k_{\!\perp})$,
corresponding to the Sawicki transformation in the 
Compendium \cite{Com00}. 
The integral equation (\ref{IntEqn}) is converted into a matrix equation
by Gauss-Legendre quadratures in the off-shell mass 
$\mu=2\vert\vec k \vert$ and the polar angle $\theta$, 
with $N_1$ and $N_2$
integration points, respectively.
The Hamiltonian matrix is subsequently diagonalized  
by numerical methods. 
In Ref.~\cite{KPW92} the fine structure constant was set to 
the very large value $\alpha=0.3$ in order to resolve the 
accumulation of the eigenvalues around $M^2\simeq (2m)^2$.
We use the same unphysically large coupling to facilitate 
the search for possible violations of rotational symmetry in the spectrum.
Because of the inherent qua\-dratically integrable
Coulomb singularity ($Q^{-2}$) convergence of the eigenvalues 
with the number of the (Gaussian) integration points is extremely slow. 
Crucial improvement is achieved by using the
Nystr{\o}m method \cite{Nystrom}. Its essence
is to add a diagonal numerical (Coulomb) counter term
and to subtract its discretized version.
In contrast to the pure Coulomb case, 
where the only counter term can be calculated 
analytically \cite{Woelz90},
the fine and hyperfine interactions in Eq.(\ref{IntEqn}) 
require several counter terms.
For $J_z{=}0$ different diagonal matrix elements occur
for parallel and anti-parallel helicities;
they require two counter terms. 
Since both amplitudes have a similar singularity structure and 
comparable values, Krautg\"artner {\it et al.} have used the
same semi-analytical counter term in both diagonal matrix elements.
Despite of this unnecessary simplification the convergence of
the eigenvalues was satisfactory.
In order to analyze the multiplet 
structure of the spectrum in all sectors of $J_z$, 
we have to improve the accuracy of this numerical algorithm. 
When $J\neq 0$, one has four different diagonal elements
in the helicity matrix, 
one of which is much smaller than the others. 
To meet these numerical requirements, we avoid analytical 
treatment of the counter terms altogether, 
and integrate the counter terms
numerically with a sufficiently high precision.
This requires a somewhat larger computing time, 
but one is rewarded by an improvement 
of a factor two in convergence.

\begin{figure} [t]  
\begin{minipage}[t]{68mm}
  \resizebox{1.0\textwidth}{!}{\includegraphics{wf_0j10.epsi}}
\caption{\label{fig:1}  
   Triplet wavefunction $\psi_{\uparrow\uparrow}$
   as a function of $x$ and $\vert\vec k_\perp\vert$, 
   at $\varphi=0$.} 
\end{minipage} \ \hfill
\begin{minipage}[t]{68mm}
  \resizebox{1.0\textwidth}{!}{\includegraphics{wf_0j11.epsi}}
\caption{\label{fig:2}  
   Triplet wavefunction $\psi_{\uparrow\downarrow}$ is
   for $J_z=1$, $\Lambda=m$, $N_1=41$, and $N_2=11$.}
\end{minipage} 
\end{figure}

The wavefunctions are generated simultaneously with the spectrum.
The two components of the lowest state ($J_z=0$)
with anti-parallel and parallel helicities, 
look, up to a scaling factor, like the wavefunctions with 
parallel and anti-parallel helicities of 
Figs.~\ref{fig:1}-\ref{fig:4}, respectively. 
The shapes and the peak values are the same as in Ref.~\cite{KPW92}.
The plots in Ref.~\cite{KPW92} seemed to indicate 
numerical problems because they showed internal structure.
We found that this is due to 
numerical mistakes in the graphing package.

\begin{figure} [t]  
\begin{minipage}[t]{68mm}
  \resizebox{1.0\textwidth}{!}{\includegraphics{wf_0j12.epsi}}
\caption{\label{fig:3}  
   Triplet wavefunction $\psi_{\downarrow\uparrow}$.}
\end{minipage} \ \hfill
\begin{minipage}[t]{68mm}
  \resizebox{1.0\textwidth}{!}{\includegraphics{wf_0j13.epsi}}
\caption{\label{fig:4}  
   Triplet wavefunction $\psi_{\downarrow\downarrow}$.}
\end{minipage} 
\end{figure}
Let us discuss briefly the properties of the $J_z{=}1$ 
wavefunctions, as displayed in Figs.~\ref{fig:1}--\ref{fig:4}.
Due to the lower symmetry, the wave functions for $J_z \neq 0$ 
show more structure than those for $J_z = 0$.
The wave functions with $J_z \neq 0$ have four components 
corresponding to the four different helicity combinations.
We can see immediately from Figs.~\ref{fig:2} and \ref{fig:3} that  
the components for anti-parallel helicities are identical.
The components for parallel helicities in Figs.~\ref{fig:1}  
and \ref{fig:4}  have rather disjunct properties: 
the ($\uparrow\uparrow$)-component peaks at 
$x = 0.5$ and $k_{\!\perp} = 0$ 
and is rotationally invariant. 
The ($\downarrow\downarrow$)-com\-po\-nent
vanishes at $k_{\!\perp} = 0$ and is shaped more like 
the components with anti-parallel helicities. 
Note the very different peak values:
the anti-parallel components are suppressed by a factor of 40
as compared to the ($\uparrow\uparrow$)-component, 
the ($\downarrow\downarrow$)-component is suppressed
by a factor of 1400!
We emphasize that the wave functions are {\em not}
rotationally invariant, despite the fact that their non-relativistic
analogue is an $s$-wave.
One reason is the fact that rotations around the transverse 
axes are dynamical. Another way of seeing this is to
transform the light-cone variables to quasi-equal-time coordinates.
The Jacobian of the transformation breaks rotational invariance due to the term
$k_z^2/(m^2 + \vec k _{\!\perp}^2 + k_z^2)$.
The breaking of rotational invariance of the wave function 
is noticeable only for large momenta $|\vec k|\gg m$, 
or, correspondingly, large cut-offs $\Lambda$. 
As an example, the triplet wave function $1^3S_1(\uparrow\uparrow)$ 
is plotted in Fig.~\ref{fig:diff-b} as a function of the off-shell mass $\mu$.
Obviously, the wave function is not isotropic but depends 
on the angle $\theta$ for sufficiently large values of the
off-shell-mass.
We note that with a cut-off of $\Lambda=m$ in place, 
the discrepancy of curves with different $\cos\theta$
would not be resolvable.

\textbf{Conclusions}:\\
(1) The counter term technology of Krautg\"artner {\it el al.} \cite{KPW92}
can be implemented without unnecessary simplifications.  \\
(2) The generalization to all values of the angular momentum projection 
$J_z$ is possible and novel.  \\
(3) The dependence on the cut-off is logarithmic,
and thus renormalizable.  \\
(4) Possible problems with rotational symmetry  
can be dealt with easier in practice than anticipated 
by more formal investigations in the literature. \\ 
(5) The eigenvalues for different $J_z$ arrange themselves
in highly degenerate multiplets.
This in turn lets us conclude that the effective interaction
of Krautg\"art\-ner {\it et al.} \cite{KPW92} is
consistent with the rotational thus Lorentz-invariance
of the Lagrangian, and this holds probably for
all light-cone based approaches.


\begin{thebibliography}{99}
\bibitem{BrodskyPauliPinsky}
   S.J. Brodsky, H.C. Pauli and S.S. Pinsky,
   Phys. Rep. \textbf{301} (1998) 299-486.
\bibitem{tHooft}
   G.'t Hooft, Nucl. Phys. \textbf{B75} (1974) 461.
\bibitem{Bassetto}
   A. Bassetto and L. Griguolo,
   Phys. Lett. \textbf{B443} (1998) 325.
\bibitem{Karmanov98}
   J. Carbonell, B. Desplanques, V.A. Karmanov and J.F. Mathiot,\\ 
   Phys. Rep. \textbf{300} (1998) 215-347.
\bibitem{Tang91}
   A. Tang, S.J. Brodsky, and H.C. Pauli,
   Phys. Rev. \textbf{D44} (1991) {1842}.
\bibitem{Kaluza92}
   M. Kalu{\v{z}}a and H.-C.~Pauli,
   Phys. Rev. \textbf{D45} (1992) 2968-2981.
\bibitem{Wilson}
   D. Mustaki, S. Pinsky, J. Shigemitsu, and K.G. Wilson,
   Phys. Rev. \textbf{D43} (1991) 3411;
   R.J. Perry, A. Harindranath, and K.G. Wilson,
   Phys. Rev. Lett. \textbf{65} (1990) 2959.
\bibitem{Jones96a}
   B.D. Jones, R.J. Perry, and S.D. G{\l}azek,
   Phys. Rev. \textbf{D55} (1997) 6561-6583.
\bibitem{KPW92}
   M. Krautg\"artner, H.C. Pauli, and F. W\"olz, 
   Phys. Rev. \textbf{45} (1992) 3755--3774.
\bibitem{Pau00}
   H.C. Pauli, this volume.
\bibitem{Leutwyler}
   H. Leutwyler and J. Stern,
   Phys. Lett. \textbf{B69} (1977) 207.
\bibitem{Hil00}
   J. Hiller, this volume.
\bibitem{MorseFeshbach}
   P.M. Morse, H. Feshbach,
   Methods in Theoretical Physics,
   McGraw-Hill 1953.
\bibitem{TamDancoff}
   I.J. Tamm, J. Phys. \textbf{9} (1945) 449;
   S.M. Dancoff, Phys. Rev. \textbf{78} (1950) 382.
\bibitem{BLepage80}
   G.P. Lepage and S.J. Brodsky, Phys. Rev. \textbf{D22} (1980) 2157.
\bibitem{Com00}
   Compendium, in the appendix to this volume.
\bibitem{TrittmannPauli97}
   U. Trittmann, Ph.D. thesis, University of Heidelberg 1996;
   hep-th/9704215.
\bibitem{TrittmannPauli97b}
   U. Trittmann and H.C. Pauli,
   hep-th/9704215, hep-th/9705021.
\bibitem{Trittmann97}
   U. Trittmann, hep-th/9705072, hep-th/9706055.
\bibitem{Hornbostel}
   K. Hornbostel, S.J. Brodsky, and H.C. Pauli,
   Phys. Rev. \textbf{D41} (1990) 3814.
\bibitem{Fermi30}
   E. Fermi,
   Z. Phys. \textbf{60} (1930) 320--333.
\bibitem{Gupta89}
   G.T. Bodwin, D.R. Yennie and C.J.~Suchyta,
   Rev. Mod. Phys. \textbf{57} (1985) 723;
   S.N. Gupta, W.W. Repko and C.J. Suchyta,
   Phys. Rev. \textbf{D40} (1989) 4100--4104.
\bibitem{Nystrom}
   W. Hackbusch, Integralgleichungen, 
   Teubner Verlag, Stuttgart 1989.
\bibitem{Woelz90}
   F. W\"olz,
   master's thesis (diploma), University of Heidelberg 1990.
\end{thebibliography}
\end{document}